\documentclass{article}

\usepackage[letterpaper,top=2cm,bottom=2cm,left=3cm,right=3cm,marginparwidth=1.75cm]{geometry}
\usepackage{amsmath}
\usepackage{graphicx}
\usepackage[colorlinks=true, allcolors=blue]{hyperref}

\newcommand{\sm}[1]{\small{#1}}
\newcommand{\mt}[1]{\mathcal{#1}}

\title{Two-electron atomic systems. A simple method for calculating the ground state near the nucleus. Some applications.}
\author{Evgeny Z. Liverts}

\begin{document}
\maketitle

\begin{abstract}
A simple method of variational calculations of the electronic structure of a two-electron atom/ion, primarily near the nucleus, is proposed.
The method as a whole consists of a standard solution of a generalized matrix eigenvalue equation, all matrix elements of which are reduced to a numerical calculation of one-dimensional integrals.
Distinctive features of the method are:
The use of the hyperspherical coordinate system.
The inclusion of logarithms of the hyperspherical radius R in the basis functions, similar to the Fock expansion.
Using a special basis function including the leading angular Fock coefficients to provide the correct behavior of the wave function near the nucleus.
The main numerical parameters characterizing the properties of the helium atom and a number of helium-like ions near the nucleus are calculated and presented in tables.
Among others, the specific coefficients $a_{21}$ of the Fock expansion, which can only be calculated using a wave function with the correct behavior near the nucleus, are presented in table and graphs.

\end{abstract}

\section{Introduction}\label{S0}

The trivial statement is that all the main features of a two-electron atomic system with an infinitely massive nucleus of the charge $Z$ and non-relativistic energy $E$ are defined by the corresponding wave function (WF) $\Phi(r_1,r_2,r_{12})$, where $r_1$ and $r_2$ are the electron-nucleus distances, and $r_{12}$ is the distance between the electrons.
It is also well known  that the behavior of the ground state WF in the vicinity of the nucleus located at the origin is determined by the Fock expansion (FE) \cite{FOCK}
\begin{equation}\label{1}
\Phi(r_1,r_2,r_{12})\equiv\Psi(R,\alpha,\theta) = \sum_{k=0}^\infty R^k\sum_{p=0}^{[k/2]}\psi_{k,p}(\alpha,\theta)\ln^p R,
\end{equation}
where the hyperspherical coordinates (HSC) $R,~\alpha$ and $\theta$ are defined by the relations
\begin{equation}\label{2}
R=\sqrt{r_1^2+r_2^2},~~~~\alpha=2\arctan\left(\frac{r_2}{r_1}\right),~~~~\theta=\arccos\left(\frac{r_1^2+r_2^2-r_{12}^2}{2 r_1 r_2}\right),
\end{equation}
whereas $\psi_{k,p}(\alpha,\theta)$ are the angular Fock coefficients (AFC), many of which have been calculated previously (see, e.g., \cite{AB1}, \cite{LEZ4}, \cite{LEZ2} and references therein).
It is worth noting that the exact representation of the AFCs $\psi_{k,p}$ with even $k$ and maximum values $p=k/2$ was recently presented in Ref. \cite{LEZ0}. These AFCs will be used in what follows.
Recall that the convergence of expansion (\ref{1}) has been proven in the work \cite{MORG}.

There are a large number of methods for calculating the electronic structure of the two-electron atomic systems.
An excellent review on this topic can be found in Refs. \cite{AB1,NAK,ROD,FOR,DRK}.
However, we are known only one technique that correctly represents the WF near the nucleus.
It is the so called correlation function hyperspherical harmonic  method (CFHHM) \cite{HM0,HM1,HM2}.
This method uses an expansion similar to the FE to represent the WF near the nucleus, and uses an expansion in hyperspherical harmonics (HHs) to represent the AFCs.
However, the HH-expansion is known to converge very slowly.
Although this method makes it possible to increase the convergence of the HH-expansion,
a sufficiently good accuracy requires a large HH basis size (thousands of BFs).
Moreover, instead of the widely used variational method for solving the problem, an algebraic method for solving coupled systems of ordinary differential equations is used.
All of the above creates great computational difficulties in implementing the method.

It follows that it would be extremely useful to develop a much simpler method for calculating the WF with correct behavior near the nucleus.
The main objective of this article is to present such a method, as well as some of its non-trivial applications.

\section{General characteristics of the method}\label{S1}

The simplicity of the proposed method primarily implies the use of a standard variational method, which involves solving the generalized matrix eigenvalue equation
\begin{equation}\label{3}
\hat{\mathcal{V}}+\hat{\mathcal{T}}=E \hat{\mathcal{S}},
\end{equation}
where $\hat{\mathcal{S}}$, $\hat{\mathcal{V}}$ and $\hat{\mathcal{T}}$ are the overlap, potential and kinetic energy matrices, respectively, the elements of which are defined as follows:
\begin{equation}\label{4}
\mathcal{S}_{j,j'}=\int{f_j f_{j'}dv},~~~~~~~~ \mathcal{V}_{j,j'}=\int{f_j V f_{j'}dv}, ~~~~~~~~\mathcal{T}_{j,j'}=\int{f_j T f_{j'}dv}.
\end{equation}
The number of basis functions (BFs) $f_j$ is limited by the basis size (BS) $N_b$ $(1\le j\le N_b)$.
The resulting WF can then be constructed in the form
\begin{equation}\label{5}
\Psi=\sum_{j=1}^{N_b} C_j f_j,
\end{equation}
where $C_j$ is the $j-th$ component of the eigenvector corresponding to solution of Eq. (\ref{3}) with eigenvalue $E$.
Note that all matrix elements (MEs) will be calculated using the HSC defined by Eq. (\ref{2})
and the \emph{atomic units} system.
This choice of coordinate system will help us to construct a WF, $\Psi$ with the behavior near the nucleus represented by the FE (\ref{1}).
Thus, for the volume element in the HSC we have \cite{AB1}:
\begin{equation}\label{6}
dv= R^5 dR~d\Omega,~~~~~~d\Omega=\pi^2 \sin^2 \alpha~d\alpha \sin \theta~d\theta,~~~~~~~~R\in[0,\infty),~~\alpha\in[0,\pi],~~\theta\in[0,\pi]
\end{equation}
The kinetic energy operator $T$ is equal to $-(1/2)\Delta$, where the Laplacian is:
\begin{equation}\label{7}
\Delta=\frac{\partial^2}{\partial R^2}+\frac{5}{R}\frac{\partial}{\partial R}-\frac{\Lambda^2}{R^2}.
\end{equation}
The hyperspherical angular momentum operator $\Lambda^2\equiv \Lambda^2(\alpha,\theta)$ is defined as follows (see, e.g., \cite{AB1}, \cite{LEZ4}):
\begin{equation}\label{8}
\Lambda^2=-4\left[\frac{\partial^2}{\partial\alpha^2}+2\cot\alpha\frac{\partial}{\partial\alpha}+ \frac{1}{\sin^2\alpha}\left( \frac{\partial^2}{\partial \theta^2}+\cot \theta \frac{\partial}{\partial\theta}\right)\right].
\end{equation}
It is reasonable to discuss the form of potential operator $V$ for a two-electron atomic system later.
At this stage, we will begin to introduce a set of BFs, which is the \emph{main feature} of the\emph{ proposed method}.
In particular, we propose to divide the entire set of BFs into two unequal parts.
The first part will consist of only one, but rather complex \underline{\emph{special}} basis function (SBF). The second part will be represented by a large set of simple BFs.

To proceed to introducing SBF, let's first recall the form of AFC for the linear in $R$ term of the FE:
\begin{equation}\label{10}
\psi_{1,0}(\alpha,\theta)=-\frac{Z(r_1+r_2)}{R}+\frac{r_{12}}{2R}=-Z \eta+\frac{1}{2}\xi,
\end{equation}
where
\begin{equation}\label{11}
\eta=\sqrt{1+\sin \alpha},~~~~~~\xi=\sqrt{1-\sin \alpha \cos \theta}.
\end{equation}
Secondly, it was already mentioned in the "Introduction" that the AFC $\psi_{k,k/2}\equiv\psi_{k,k/2}(\alpha,\theta)$ with even $k$ have recently been presented in the following closed analytical form:
\begin{equation}\label{13}
\psi_{k,k/2}=Z^{\frac{k}{2}}\sum_{l=k/2,(-2)}^0 \mathbf{c}_{k,l}Y_{k,l}(\alpha,\theta),
\end{equation}
where $Y_{k,l}(\alpha,\theta)$ are HHs, and the closed form of the coefficients $\mathbf{c}_{k,l}$ for $k\leq10$ can be found in Ref. \cite{LEZ0}.
It is seen that both the AFCs (\ref{13}) and the linear AFC (\ref{10}) depend on the nucleus charge $Z$ and do not depend on the energy $E$. However, more importantly, the right-hand side (RHS) of the representation (\ref{13}) can be reduced to the polynomial in variables $\eta$ and $\xi$ defined by Eq.(\ref{11}). In particular, we obtain:
\begin{equation}\label{15}
\psi_{k,k/2}=\left(\frac{Z}{\pi}\right)^{\frac{k}{2}}
 \widetilde{b}_{\frac{k}{2}}\sum_{n,m} b_{\frac{k}{2},n,m}\eta^n\xi^m,
\end{equation}
where the coefficients $\widetilde{b}_{\frac{k}{2}}$ and $b_{\frac{k}{2},n,m}$ are represented in Table \ref{T1}.

Taking into account Eqs. (\ref{10}) and (\ref{15}), and also the fact that the products $(\ln R)^{k/2} R^k\psi_{k,k/2}$ represent the leading terms of the logarithmic series \cite{LEZ0} of the FE, it is reasonable to introduce the SBF in the form:
\begin{equation}\label{16}
f_0\equiv f_0(R,\alpha,\theta)=\exp\left[-R\left(Z \eta-\frac{1}{2}\xi\right)\right]\sum_{k'=0(2)}^K R^{k'}(\ln R)^{k'/2}\psi_{k',k'/2}(\alpha,\theta),
\end{equation}
where the AFCs are represented by Eq.(\ref{15}), and $K$ is the parameter of the problem under consideration.
It should be emphasized that the SBF (\ref{16}) is designed to serve two purposes. The first one is to provide the correct representation of the FE terms with even $k$ and maximum power $p=k/2$ of $\ln R$. And the second one is to guarantee the correct form of the AFC $\psi_{1,0}(\alpha,\theta)$ or (which is equivalent) to guarantee that the WF will satisfy the Kato's cusp conditions \cite{KAT,MYE}.

The main part of the complete basis set (BS) can be taken as:
\begin{equation}\label{17}
f_{kp,nm}\equiv f_{kp,nm}(R,\alpha,\theta)=\exp\left[-R\left(\lambda \eta-\gamma\xi\right)\right]R^k(\ln R)^p \eta^n\xi^m,~~~~~~~~~~~~~~~~~~(k>1)
\end{equation}
where, in general, $\lambda$ and $\gamma$ are variational parameters.
It is clear that the BS (\ref{17}) is bounded by $k\geq 2$, since the constant ($k=0$) and linear ($k=1$) terms in $R$ are described by the SBF (\ref{16}).

The Coulomb potential operator in the HSC can now be introduced as follows:
\begin{equation}\label{18}
V\equiv\frac{1}{r_{12}}-Z\left(\frac{1}{r_1}+\frac{1}{r_2}\right)=\frac{1}{R}\left(\frac{1}{\xi}-\frac{2 Z \eta}{\sin \alpha}\right).
\end{equation}
Note that according to our building of the BS, the MEs (\ref{4}) of each of the three matrices $\hat{\mathcal{T}}$, $\hat{\mathcal{V}}$ and $\hat{\mathcal{S}}$ can be divided into three groups.
The first group includes MEs between BFs of the form (\ref{17}), the second group - between BFs (\ref{17}) and (\ref{16}), and the third group - between BFs (\ref{16}).

It is also important to note that the constant $\pi^2$ included in the angular space element $d\Omega$ (see definitions (\ref{6})) is a factor for all terms of the matrix equation (3), and will therefore be omitted in what follows.

\section{Overlap and potential energy matrices}\label{S2}

According to definition (\ref{4}), the overlap MEs calculated between BFs of the form (\ref{17}) are:
\begin{eqnarray}\label{20}
\mathcal{S}_{kp,nm;k'p',n'm'}\equiv \pi^{-2}\int{f_{kp,nm} f_{k'p',n'm'}dv}=
~~~~~~~~~~~~~~~~~~~~~~~~~~~~~~~~~~~~\nonumber~\\
\int_0^\infty R^5 d R\int_0^\pi\sin^2 \alpha~d\alpha\int_0^\pi
 R^{k+k'}(\ln R)^{p+p'}
 \exp\left[-2 R(\lambda \eta-\gamma \xi) \right]\eta^{n+n'}\xi^{m+m'}\sin \theta~ d \theta=
 ~\nonumber~~~~\\
 \mathcal{K}_2(k+k',p+p',n+n',m+m';2\lambda,2\gamma),
~~~~~~~~~~~~~~~~~~~~~
\end{eqnarray}
where the specific calculations of the general 3D (three-dimensional) integral
\begin{equation}\label{21}
\mathcal{K}_L(k,p,n,m;a,b)=
\int_0^\pi\eta^n\sin^L \alpha~d\alpha\int_0^\pi
 \xi^m\sin \theta~ d \theta\int_0^\infty R^{k+5}(\ln R)^p\exp\left[-R(a \eta-b \xi) \right] d R~~~~
\end{equation}
will be discussed in the Appendix.


The overlap MEs calculated between BFs (\ref{17}) and SBF (\ref{16})
can be represented in terms of the $\mathcal{K}$-integrals (\ref{21}) as follows:
\begin{equation}\label{22}
\mathcal{S}_{kp,nm;0}\equiv \pi^{-2}\int{f_{kp,nm} f_0 dv}=
\sum_{k'=0,(2)}^{K}
\left(\frac{Z}{\pi}\right)^{\frac{k'}{2}}
 \widetilde{b}_{\frac{k'}{2}}\sum_{n',m'} b_{\frac{k'}{2},n',m'}
 \mathcal{K}_2\left(k_1,p_2,n_1,m_1;\lambda+Z,\gamma+\frac{1}{2}\right),
\end{equation}
where we introduced notations:
\begin{equation}\label{23}
k_1=k+k',~~~~~~~p_1=p+p',~~~~~~~p_2=p+k'/2,~~~~~~~n_1=n+n',~~~~~~~m_1=m+m',~~~~
\end{equation}
that remain relevant in what follows.

It is worth noting that the integer parameter $K$ is determined by the number of the FE terms $R^{k'} (\ln R)^{p'}\psi_{k',p'}(\alpha,\theta)$ with the maximum value $p'=k'/2$ included into the SBF (\ref{16}) for this method option.

The third group of MEs, consisting of only one overlap ME calculated between SBFs (\ref{16}), can be determined as:
\begin{eqnarray}\label{24}
\mathcal{S}_{0;0}\equiv \pi^{-2}\int{f_0^2 dv}=
~~~~~~~~\nonumber~~~~~~~~~~~~~~~~~~~~~~~~~~~~~~~~~~~~~~~~~~~~~~~~~~~~~~~\\
\sum_{k=0,(2)}^K
\widetilde{b}_{\frac{k}{2}}\sum_{n,m} b_{\frac{k}{2},n,m}
\sum_{k'=0,(2)}^{K}
\left(\frac{Z}{\pi}\right)^{\frac{k+k'}{2}}
 \widetilde{b}_{\frac{k'}{2}}\sum_{n',m'} b_{\frac{k'}{2},n',m'}
 \mathcal{K}_2\left(k_1,\frac{k_1}{2},n_1,m_1;2Z,1\right).
\end{eqnarray}


Following the previously proposed scheme for calculating the overlap matrix, it is easy to derive the corresponding formulas for the elements of the potential energy matrix.
Thus, using the notations (\ref{23}), we obtain:
\begin{eqnarray}\label{25}
\mathcal{V}_{kp,nm;k'p',n'm'}\equiv \pi^{-2}\int{f_{kp,nm}
\frac{1}{R}\left(\frac{1}{\xi}-\frac{2 Z \eta}{\sin \alpha}\right) f_{k'p',n'm'}dv}=
 ~\nonumber~~~~~~~~~~~~~~~~~~~~~~~~~~\\
\mathcal{K}_2(k_1-1,p_1,n_1,m_1-1;2\lambda,2\gamma)
 -2Z\mathcal{K}_1(k_1-1,p_1,n_1+1,m_1;2\lambda,2\gamma) ,
\end{eqnarray}
\begin{eqnarray}\label{26}
\mathcal{V}_{kp,nm;0}\equiv \pi^{-2}\int{f_{kp,nm}
\frac{1}{R}\left(\frac{1}{\xi}-\frac{2 Z \eta}{\sin \alpha}\right) f_0 dv}=
\sum_{k'=0,(2)}^K
\left(\frac{Z}{\pi}\right)^{\frac{k'}{2}}
 \widetilde{b}_{\frac{k'}{2}}\sum_{n',m'} b_{\frac{k'}{2},n',m'}\times
 ~~~~~~~~~~~~~~~~\nonumber~~~~~\\
 \left[\mathcal{K}_2\left(k_1-1,p_2,n_1,m_1-1;\lambda+Z,\gamma+\frac{1}{2}\right)-
 2Z\mathcal{K}_1\left(k_1-1,p_2,n_1+1,m_1;\lambda+Z,\gamma+\frac{1}{2}\right)
 \right],~~~~~~~~~~
\end{eqnarray}
\begin{eqnarray}\label{27}
\mathcal{V}_{0;0}\equiv \pi^{-2}\int{f_0^2 \frac{1}{R}\left(\frac{1}{\xi}-\frac{2 Z \eta}{\sin \alpha}\right)dv}=
\sum_{k=0,(2)}^K
\widetilde{b}_{\frac{k}{2}}\sum_{n,m} b_{\frac{k}{2},n,m}
\sum_{k'=0,(2)}^{K}
\left(\frac{Z}{\pi}\right)^{\frac{k+k'}{2}}
 \widetilde{b}_{\frac{k'}{2}}\sum_{n',m'} b_{\frac{k'}{2},n',m'}\times
 ~\nonumber~\\
 \left[\mathcal{K}_2\left(k_1-1,\frac{k_1}{2},n_1,m_1-1;2Z,1\right)
 -2Z\mathcal{K}_1\left(k_1-1,\frac{k_1}{2},n_1+1,m_1;2Z,1\right)\right].~~~~
\end{eqnarray}

\section{Kinetic energy matrix}\label{S3}

To simplify the derivation of calculation formulas for the kinetic energy MEs defined by Eqs.(\ref{4}) and (\ref{7}), we divide them into two parts as follows:
\begin{equation}\label{30}
\mathcal{T}_{j,j'}=
-\frac{1}{2}\left[\mathcal{T}_{j,j'}^{(1)}-\mathcal{T}_{j,j'}^{(2)}\right],
\end{equation}
where
\begin{equation}\label{31}
\mathcal{T}_{j,j'}^{(1)}=\pi^{-2}\int{f_{j'}
\left(\frac{\partial^2}{\partial R^2}+\frac{5}{R}\frac{\partial}{\partial R}\right) f_j dv},~~~~~~~~~~~~~~~~~
\mathcal{T}_{j,j'}^{(2)}=\pi^{-2}\int{f_{j'}\frac{\Lambda^2}{R^2} f_{j}dv}.~~~~~~~~~~~~~~
\end{equation}
To proceed, we use the results of the action of the Laplace operator components on the constituents of our BS defined by Eqs.(\ref{16}) and (\ref{17}). In particular, we obtain:
\begin{eqnarray}\label{32}
 \left(\frac{\partial^2}{\partial R^2}+\frac{5}{R}\frac{\partial}{\partial R}\right)\exp\left[-R\left(a \eta-b\xi\right)\right]R^k(\ln R)^p =
\exp\left[-R\left(a \eta-b\xi\right)\right]R^k(\ln R)^p\times
~\nonumber~\\
\left[
\frac{p(p-1)}{R^2 \ln^2 R}+\frac{2p(k+2)}{R^2\ln R}+\frac{k(k+4)}{R^2}+(a \eta-b\xi)^2
-\left(\frac{2p}{R \ln R}+\frac{2k+5}{R}\right)(a \eta-b\xi)
\right],
\end{eqnarray}
\begin{eqnarray}\label{33}
\Lambda^2\exp\left[-R\left(a \eta-b\xi\right)\right]\eta^n\xi^m =
\exp\left[-R\left(a \eta-b\xi\right)\right]\eta^n\xi^m\times
 ~~~~~~~~~~~~~~~~~~~~~~~~~~~~~~~~~~~~~~~~~\nonumber~\\
\left\{
 (m+n)(m+n+4)-2(a^2+b^2)R^2-\frac{2n(n-1)}{\eta^2}+\frac{4na R}{\eta}+a^2\eta^2R^2
 -\frac{2m(m+n+1)}{\xi^2}
\right.
~\nonumber~\\
-\frac{2(2m+n+2)b R}{\xi}+b(2m+2n+5)\xi R+b^2\xi^2R^2-a(2m+2n+1)\eta R+\frac{2ma\eta R}{\xi^2}
~\nonumber~\\
+2ab\eta \left(\frac{1}{\xi}-\xi\right)R^2+\frac{2}{\sin \alpha}
\left[-n(m+2)+\frac{m n}{\xi^2}+nb\left(\frac{1}{\xi}-\xi\right)R
\right.
~\nonumber~\\
\left.\left.
+a(m+4)\eta R-2a\eta^3 R-\frac{ma\eta R}{\xi^2}-ab\eta\left(\frac{1}{\xi}-\xi\right)R^2
\right]\right\}.
\end{eqnarray}
It can be seen that the angular parts of both results represented by the RHSs of the last two equations are expressed in terms of the angular variables $\eta$ and $\xi$, and the factor $\sin^{-1} \alpha$. This enables us to represent the kinetic energy MEs through the integrals of the form (\ref{21}).

Thus, using Eq.(\ref{31}), as well as Eqs.(\ref{32}) and (\ref{33}) with $a=\lambda$ and $b=\gamma$, we obtain for the kinetic energy MEs calculated between BFs of the form (\ref{17}):
\begin{eqnarray}\label{34}
\mt{T}_{k'p',n'm';kp,nm}^{(1)}=p(p-1)\mt{K}_2(k_1-2,p_1-2,n_1,m_1)+2p(k+2)\mt{K}_2(k_1-2,p_1-1,n_1,m_1)~~~~
~\nonumber~\\
+k(k+4)\mt{K}_2(k_1-2,p_1,n_1,m_1)+\lambda^2\mt{K}_2(k_1,p_1,n_1+2,m_1)-2\lambda\gamma\mt{K}_2(k_1,p_1,n_1+1,m_1+1)~~~~~~~~~
~\nonumber~\\
+\gamma^2\mt{K}_2(k_1,p_1,n_1,m_1+2)-2p\left[\lambda\mt{K}_2(k_1-1,p_1-1,n_1+1,m_1)-\gamma\mt{K}_2(k_1-1,p_1-1,n_1,m_1+1)\right]
~\nonumber~\\
-(2k+5)\left[\lambda\mt{K}_2(k_1-1,p_1,n_1+1,m_1)-\gamma\mt{K}_2(k_1-1,p_1,n_1,m_1+1)\right],~~~~~~~~~~~
\end{eqnarray}

\begin{eqnarray}\label{35}
\mt{T}_{k'p',n'm';kp,nm}^{(2)}=(m+n)(m+n+4)\mt{K}_2(k_1-2,p_1,n_1,m_1)-2(\lambda^2+\gamma^2)\mt{K}_2(k_1,p_1,n_1,m_1)
~\nonumber~\\
-2n(n-1)\mt{K}_2(k_1-2,p_1,n_1-2,m_1)+4 n \lambda\mt{K}_2(k_1-1,p_1,n_1-1,m_1)+\lambda^2\mt{K}_2(k_1,p_1,n_1+2,m_1)
~\nonumber~\\
-2m(m+n+1)\mt{K}_2(k_1-2,p_1,n_1,m_1-2)-2(2m+n+2)\gamma\mt{K}_2(k_1-1,p_1,n_1,m_1-1)
~\nonumber~\\
+\gamma(2m+2n+5)\mt{K}_2(k_1-1,p_1,n_1,m_1+1)+\gamma^2\mt{K}_2(k_1,p_1,n_1,m_1+2)
~~~\nonumber~\\
-\lambda(2m+2n+1)\mt{K}_2(k_1-1,p_1,n_1+1,m_1)+2m\lambda\mt{K}_2(k_1-1,p_1,n_1+1,m_1-2)
~\nonumber~\\
+2\lambda\gamma\left[\mt{K}_2(k_1,p_1,n_1+1,m_1-1)-\mt{K}_2(k_1,p_1,n_1+1,m_1+1)\right]~~~~~
~~~~~~\nonumber~~~\\
-2n(m+2)\mt{K}_1(k_1-2,p_1,n_1,m_1)+2mn\mt{K}_1(k_1-2,p_1,n_1,m_1-2)
~~~~~~\nonumber~\\
+2n\gamma\left[\mt{K}_1(k_1-1,p_1,n_1,m_1-1)-\mt{K}_1(k_1-1,p_1,n_1,m_1+1)\right]+2\lambda(m+4)\mt{K}_1(k_1-1,p_1,n_1+1,m_1)
~\nonumber~\\
-4\lambda\mt{K}_1(k_1-1,p_1,n_1+3,m_1)-2m\lambda\mt{K}_1(k_1-1,p_1,n_1+1,m_1-2)~~~~~
~\nonumber~\\
-2\lambda\gamma\left[\mt{K}_1(k_1,p_1,n_1+1,m_1-1)-\mt{K}_1(k_1,p_1,n_1+1,m_1+1)\right].~~~~~~~~~~~~~
\end{eqnarray}
For simplicity, we replaced the designations $\mt{K}_L(k,p,n,m;2\lambda,2\gamma)$ with $\mt{K}_L(k,p,n,m)$ for all $\mt{K}$-integrals in Eqs.(\ref{34}) and (\ref{35}), that is, we simply omitted the last two \emph{common} parameters.
We also used the notations introduced in Eq.(\ref{23}).

Similarly, for the kinetic energy MEs calculated between BFs (\ref{16}) and SBF (\ref{17}), we obtain:
\begin{equation}\label{36}
\mt{T}_{0;kp,nm}^{(j)}=\sum_{k'=0,(2)}^{K}
\left(\frac{Z}{\pi}\right)^{\frac{k'}{2}}
 \widetilde{b}_{\frac{k'}{2}}\sum_{n',m'} b_{\frac{k'}{2},n',m'}
 \widehat{\mt{T}}_{k'p',n'm';kp,nm}^{(j)},~~~~~~~~~~~~~~(j=1,2)
\end{equation}
where $\widehat{\mt{T}}_{k'p',n'm';kp,nm}^{(1)}$ and $\widehat{\mt{T}}_{k'p',n'm';kp,nm}^{(2)}$ represent the RHSs of Eqs.(\ref{34}) and (\ref{35}), respectively, but in which $p_1$ should be replaced by $p_2$ (see Eq.(\ref{23})), and the last two common parameters in the $\mt{K}$-integrals should be taken successively as $\lambda+Z$ and $\gamma+1/2$, similar to Eqs.(\ref{22}) and (\ref{26}).

Accordingly, the kinetic energy MEs calculated between SBFs can be represented as:
\begin{equation}\label{37}
\mt{T}_{0;0}^{(j)}=\sum_{k=0,(2)}^K
\widetilde{b}_{\frac{k}{2}}\sum_{n,m} b_{\frac{k}{2},n,m}
\sum_{k'=0,(2)}^{K}
\left(\frac{Z}{\pi}\right)^{\frac{k+k'}{2}}
 \widetilde{b}_{\frac{k'}{2}}\sum_{n',m'} b_{\frac{k'}{2},n',m'}\widehat{T}_{k'p',n'm';kp,nm}^{(j)},~~~~~~~(j=1,2)
\end{equation}
where $\widehat{T}_{k'p',n'm';kp,nm}^{(1)}$ and $\widehat{T}_{k'p',n'm';kp,nm}^{(2)}$ represent the RHSs of Eqs.(\ref{34}) and (\ref{35}), respectively, but in which the parameters $p_1,\lambda$ and $\gamma$ should be replaced by $k_1/2,Z$ and $1/2$, respectively, and the last two common parameters in the corresponding $\mt{K}$-integrals should be taken successively as $2Z$ and $1$.

\section{Numerical results and discussion}\label{S4}

For convenience, we will call the method described above as the variational method of near the nucleus calculations (VMNN).
We applied VMNN to calculate the ground state energies $E$ and the corresponding WFs $\Psi$ for the helium atom and a number of two-electron ions.

The results corresponding to the simplest version of the method with $\lambda=Z$ and $\gamma=1/2$ are presented in Table \ref{T2}.
Note that we call this variant "simplest" because the mentioned choice of parameters implies that both SBF and the set of BFs (\ref{17}) contain the same exponential.
In the sixth column we present one of the most important characteristics of the obtained state, namely, the corresponding deviation from the \emph{virial} theorem for the Coulomb interactions. The $10^{-M}$ factor has been replaced with $(-M)$.
To confirm the main purpose of VMNN, in the last two columns we present two matrix elements of the relevant operators, which represent the main properties of WF near the nucleus and can be calculated in the following simple way:
\begin{equation}\label{41}
\langle\delta(\textbf{r}_1)\rangle=\frac{4\pi}{N}\int_0^\infty \Psi^2(R,0,0)R^2 dR,~~~~
\langle\delta(\textbf{r}_1)\delta(\textbf{r}_2)\rangle=\frac{1}{N}\Psi^2(0,0,0),~~~~
\end{equation}
where $\delta(\textbf{r})$ denotes the Dirac delta function, and the normalization coefficient is:
\begin{equation}\label{42}
N=\int \Psi^2(R,\alpha,\theta)dv.
\end{equation}
The last significant digit in energy and matrix elements (\ref{41}) is the first one that differs from the more accurate calculations (see, e.g., \cite{FR0,FR1,FR2,DRK}) for $2\leq Z\leq28$, or otherwise, from the Pekeris-like method (PLM) \cite{LEZ1,LEZ1a} for $Z\geq 30$.

It is well-known that the negative ion of hydrogen is the only ion in the two-electron atomic sequence that has only one bound (ground) state. In other words, there are no excited bound states in this ion.
At the same time, its WF is rather diffuse.
These features cause special computational difficulties when using not only variational methods, including the one presented here or PLM, but also CFHHM. To solve this problem, significantly higher calculation accuracy and a significantly longer basis size are required.
It should be emphasized that the given problem concerns only a two-electron negative ion with a nuclear charge $Z$ equal to exactly $1$.
Note that there is no mathematical (computational) problem of calculating the electronic structure for a two-electron system with non-integer $Z$. In particular, the unrealistic case $Z=1.5$
presented in Table \ref{T2}, was calculated without any specific problems.
These results will be useful in solving the problem that will be discussed next.

It is seen from the representation (\ref{5}) that the WF is defined, first of all, by the set of BFs, which in turn, are determined by the sets of integers $\left\{ k,p,n,m \right\}$ for the BFs (\ref{17}), and by integer number $K$ for the SBF (\ref{16}).
To construct the large set (\ref{17}) we used two characteristic numbers $k_{max}$ and $\omega_{max}$, such that
\begin{equation}\label{45}
1<k\leq k_{max},~~~~~~~~~~~0\leq n\leq \omega_{max},~~~~~~~~~~~~0\leq m\leq \omega_{max},
\end{equation}
but under additional condition
\begin{equation}\label{46}
k+p+n+m \leq \max(k_{max},\omega_{max}).
\end{equation}
The power $p$ of logarithm can vary as follows: $0\leq p \leq [k/2] $, where $[x]$ denotes (as in the FE) the integer part of $x$.
It is clear that all cases $p=k/2$ for even $k$ must be excluded from the set (\ref{17}), since these cases are represented by the SBF (\ref{16}) with upper limit $K=2[k_{max}/2]$.
To reduce the length of the set (\ref{17}), one can use the well-known closed form of the AFC \cite{AB1,LEZ4}
\begin{equation}\label{47}
\psi_{3,1}(\alpha,\theta)=\frac{Z(\pi-2)}{36\pi}\left[6Z \eta(1-\xi^2)+\xi(5\xi^2-6)\right].
\end{equation}
This enables us to restrict BFs (\ref{17}) with $\left\{k,p\right\}=\left\{3,1\right\}$ to angular parts defined by only four sets of integers: $\left\{n,m\right\}=\left\{1,0\right\},\left\{1,2\right\},\left\{0,1\right\},\left\{0,3\right\}$.

Thus, using all the above conditions, we obtain BSs, in particular those presented in Table \ref{T2}, of length $517$ and $442$ for
$\left\{k_{max},\omega_{max} \right\}$ equal to $\left\{7,13\right\}$ and $\left\{8,12\right\}$, respectively.

The following comments are needed regarding the FE parameter $a_{21}$ calculated by VMNN and presented in Table \ref{T2}.

The AFCs $\psi_{k,p}$ satisfy the Fock recurrence relation (FRR)
\begin{equation}\label{50}
\left[ \Lambda^2-k(k+4)\right]\psi_{k,p}(\alpha,\theta)=h_{k,p}(\alpha,\theta),
\end{equation}
where the operator $\Lambda^2$ is defined by Eq.(\ref{8}).
The specific form of the RHS $h_{k,p}(\alpha,\theta)$, which is not important for our consideration, can be found, for example, in Ref. \cite{AB1}.
What really matters is that the general solution of the inhomogeneous differential equation (\ref{50}) can be expressed as the sum of a particular solution $\widetilde{\psi}_{k,p}$ of that equation and the general solution $\phi_{k,p}$ of the associated homogeneous equation
\begin{equation}\label{51}
\left[ \Lambda^2-k(k+4)\right]\phi_{k,p}(\alpha,\theta)=0,
\end{equation}
which can be represented as (see, e.g., \cite{AB1}, \cite{LEZ4})
\begin{equation}\label{52}
\phi_{k,p}(\alpha,\theta)=\sum_{l=0}^{[k/2]}a_{kl}^{(p)}Y_{kl}(\alpha,\theta),
\end{equation}
where $Y_{kl}$ are the \emph{unnormalized} HHs.
A particular solution $\widetilde{\psi}_{k,p}$ can be obtained by directly solving Eq.(\ref{50}) under suitable boundary conditions. The contribution of $\phi_{k,p}$ is determined by the coefficients $a_{k,l}^{(p)}$, which can only be found
using the WF, $\Psi(R,\alpha,\theta)$ calculated, e.g., by VMNN.

We are interested in the case $k=2,~~p=0$ which has been discussed, e.g., in  Ref. \cite {MYE}.
For this case Eq.(\ref{52}) becomes
\begin{equation}\label{53}
\phi_{2,0}(\alpha,\theta)=a_{20}^{(0)}Y_{20}(\alpha,\theta)+a_{21}^{(0)}Y_{21}(\alpha,\theta).
\end{equation}
For singlet $S$ states (including the ground state) the coefficient $a_{20}^{(0)}$ is identically zero, since the unnormalized HH, $Y_{20}(\alpha,\theta)\equiv 2\cos \alpha$ does not preserve sign under the transformation
 $\alpha \rightleftharpoons \pi-\alpha$ which essentially corresponds to a permutation of electrons.
Recall that the WF of a two-electron atomic system must preserve its parity with such a permutation.
To calculate the only remaining non-zero coefficient $a_{21}\equiv a_{21}^{(0)}$, we must first obtain the angular function $\widetilde{\phi}_{20}(\alpha,\theta)\simeq \phi_{20}(\alpha,\theta)$, which is a factor for $R^2$ in the series expansion of the WF, $\Psi(R,\alpha,\theta)$ near the the nucleus $(R\rightarrow 0)$, calculated by VMNN.
These function can then be used to calculate the required coefficient $a_{21}$ as follows:
\begin{equation}\label{55}
a_{21}=\pi^2 N_{21}^2 \int _0^{\pi}\int_0^{\pi}\widetilde{\phi}_{20}(\alpha,\theta)Y_{21}(\alpha,\theta)\sin^2\alpha\sin \theta~d\alpha~d \theta,
\end{equation}
where $N_{21}=2\pi^{-3/2}$ is the normalization coefficient for the HH, $Y_{21}(\alpha,\theta)=\sin \alpha \cos \theta$.
Note that the last relation represents a particular case for calculating the expansion coefficients in HHs for the general function of angles $\alpha$ and $\theta$ (see, e.g., \cite{AB1}).
It is the coefficients $a_{21}$, computed according to Eq.(\ref{55}), that are presented in Table \ref{T2}.

It is worth noting the following. The term $\phi_{20}(\alpha,\theta)=a_{21}\sin\alpha \cos\theta$ represents the total contribution of the HH, $Y_{21}(\alpha,\theta)$ into the AFC $\psi_{2,0}(\alpha,\theta)=\widehat{\psi}_{2,0}(\alpha,\theta)+\phi_{20}(\alpha,\theta)$.
On the other hand, in practice, all obtained particular solutions $\widehat{\psi}_{2,0}(\alpha,\theta)$ of the corresponding FRR also contain an admixture of $Y_{21}(\alpha,\theta)$.
Therefore, before adding the components  $\phi_{20}(\alpha,\theta)$ and $\widehat{\psi}_{2,0}(\alpha,\theta)$,
it is necessary to get rid of the admixture $C_{21}Y_{21}( \alpha, \theta)$ in the latter component to obtain the correct AFC, $\psi_{20}(\alpha,\theta)$.
In particular, the authors of the work \cite{AB3} were the first to obtain a particular solution $\widehat{\psi}_{2,0}(\alpha,\theta)$ in a closed analytic form (see also, \cite{FOR} and \cite{LEZ4}) for the helium-like sequence. The corresponding admixture coefficient (at least for singlet S states) was obtained \cite{LEZ3} in the form:
\begin{equation}\label{56}
C_{21}=\frac{Z\left(62+17 \pi-48 G\right)}{72\pi},
\end{equation}
where $G$ is Catalan's constant.

\begin{figure}
\caption{The Fock expansion coefficient $a_{21}$ as a function of the nucleus charge $Z$ is presented in the top row, and the accuracy estimate of the analytic interpolation function $g(Z)$ defined by equations (\ref{57})-(\ref{59}) is shown in the bottom row of the graph.}
\includegraphics[width=17.0 cm]{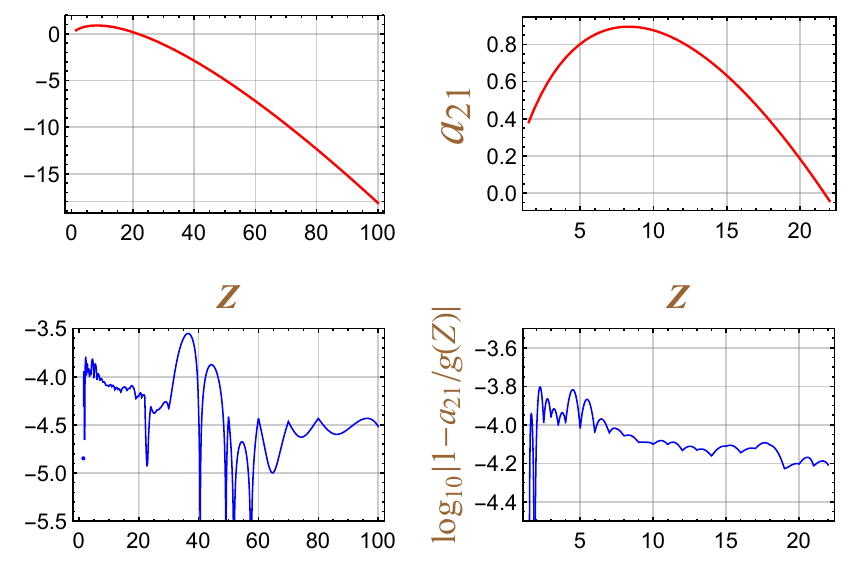}
\label{F1}
\end{figure}

The results for $a_{21}$ as a function of the nucleus charge $Z$ are graphically presented in Fig. \ref{F1}.
The curves corresponding to the full range $Z\in[1.5,100]$ and the partial range $Z\in[1.5,22]$ are shown in the left and right columns, respectively. It is seen that the function $a_{21}(Z)$ has two characteristic points. These are the maximum point $Z_{max}\simeq 8.30613237$ and the zero point $Z_0\simeq21.645108$ of the function.
Note that to obtain more accurate results, we calculated $a_{21}$ for a set of additional half-integer points, in  particular for $Z=2.5,3.5,7.5,8.5,9.5$.
It was found that a simple analytic function of the form
\begin{equation}\label{57}
g(Z)=\frac{c_{-1}}{Z}+c_0+c_1 Z(c_2+\ln Z),
\end{equation}
with properly chosen parameters $c_{-1},c_0,c_1$ and $c_2$ provides an excellent interpolation of the curve calculated using VMNN.
Direct numerical interpolation gives a good hint for the following choice of two of the four parameters:
\begin{equation}\label{58}
c_0=-\frac{1}{9},~~~~~~~~~~c_1=\frac{2-\pi}{3\pi}.
\end{equation}
It is easy to verify that the second parameter in last equation represents the factor for the AFC $\psi_{21}$.
The two remaining parameters can be found by solving the system of two equations: $g'(Z_{max})=0$ and $g(Z_0)=0$ which, given the results (\ref{58}), yield
\begin{equation}\label{59}
c_{-1}=0.0012024376,~~~~~~~~~~c_2=-3.117138.
\end{equation}
The accuracy of the resulting function $g(Z)$ was estimated using the function $\log_{10}\left|1-a_{21}/g(Z)\right|$, which characterizes the relative difference between the functions under consideration.
This evaluation function is shown in the bottom row of Fig. \ref{F1}.
It can be seen that the difference is less than a hundredth of a percent.
It is important to note that all the coefficients $a_{21}$ that were used to construct Fig. \ref{F1} and most of which are presented in Table. \ref{T2}, coincide (at least to 5 significant digits) with those calculated
using the AFCs computed by CFHHM.

It should be emphasized that although the above results confirm the correct behavior of the WF near the nucleus, this does not mean that the WF behaves incorrectly at sufficiently large $R$. On the contrary, the sufficiently high accuracy of the calculations of the energy and parameters characterizing the \emph{virial} theorem tells us the opposite.

\section{Acknowledgments}\label{S5}

I would like to thank Dr. R. Krivec for providing the AFCs calculated by CFHHM and intended to confirm the correctness of the $a_{21}$ coefficients calculated using VMNN described in this article.

\appendix
\numberwithin{equation}{section}
\section{Appendix}\label{S6}

It was shown in Sections \ref{S2} and \ref{S3} that all matrix elements needed to solve the matrix equation (\ref{3}) can be expressed in term of 3D integrals of the form (\ref{21}).
The purpose of this section is to show how the 3D integrals under consideration can be reduced to 1D integrals, which leads to increased accuracy and a reduction in computation time by hundreds of times.

The first step is to integrate over $R$ using the relation:
\begin{equation}\label{A1}
\int_0^\infty R^{k+5}(\ln R)^p \exp(-\beta R)dR=\beta^{-k-6}\sum_{t=0}^p {p\choose t}C_{p-t}^{(k)}(-\ln \beta)^t,
\end{equation}
where the coefficients
\begin{equation}\label{A2}
C_s^{(k)}=\frac{d^s \Gamma(k+6)}{dk^s}
\end{equation}
are expressed in terms of the derivatives of the gamma function.

Substituting Eq.(\ref{A1}) with $\beta=a \eta-b \xi$ into definition (\ref{21}), we obtain:
\begin{equation}\label{A3}
\mathcal{K}_L(k,p,n,m;a,b)=
2\int_0^{\pi/2} \eta^n (\sin\alpha)^L\sum_{t=0}^p(-1)^t{p\choose t}C_{p-t}^{(k)}~J_t(\alpha)~d\alpha,
\end{equation}
where
\begin{equation}\label{A4}
 J_t(\alpha)=\int_0^\pi \frac{\xi^m\ln^t(a \eta-b \xi)}{(a\eta-b \xi)^{k+6}}\sin \theta d\theta=
\frac{2}{\sin \alpha}\int_\zeta^\eta \frac{x^{m+1}\ln^t(a \eta-b x)}{(a \eta-b x)^{k+6}}dx.
\end{equation}
Function $\eta\equiv\eta(\alpha)$ is defined by Eq.(\ref{11}), and
\begin{equation}\label{A5}
 \zeta\equiv\zeta(\alpha)=\sqrt{1-\sin \alpha}.
\end{equation}
When deriving (\ref{A3}), it was taken into account that the $\alpha$-dependence of the integrand on the RHS of this equation is represented only by $\sin \alpha$, which determines the symmetry with respect to the point $\pi/2$.

The next step is to represent the integral (\ref{A4}) as:
\begin{equation}\label{A6}
 J_t(\alpha)=\frac{2}{\sin \alpha}\left[\mathcal{P}_{k+6,m+1,t}(a,b,\alpha,\eta)-\mathcal{P}_{k+6,m+1,t}(a,b,\alpha,\zeta)\right],
\end{equation}
where
\begin{equation}\label{A7}
\mathcal{P}_{k,m,t}(a,b,\alpha,\rho)=
\int_0^\rho \frac{x^{m}\ln^t(a \eta-b x)}{(a \eta-b x)^{k}}dx.
\end{equation}
The representation (\ref{A6})-(\ref{A7}) enables us to apply a general relation (see, e.g., (2.6.12.8)\cite{PRU})
\begin{eqnarray}\label{A8}
\int_0^a\frac{x^{\mu-1}(a-x)^{\nu-1}\ln^n(cx+d)}{(cx+d)^r}dx=
~~~~~~~~~~~~~~~~~~~~~~~~~~~~~~~~~~~~~~\nonumber~~~~~~~~~~~~~~~~~~~~~~~~~~~\\
a^{\mu+\nu-1}B(\mu,\nu)d^{-r}\sum_{q=0}^n(-1)^{q+n}{n\choose q}(\ln d)^q \frac{\partial^{n-q}}{\partial r^{n-q}}
~_2F_1\left(r,\mu;\mu+\nu;-\frac{a c}{d}\right),~~~~[a,\textrm{Re}~\mu,\textrm{Re}~\nu>0]
\end{eqnarray}
to obtain the closed analytic form for the integral (\ref{A7}).
$B(\mu,\nu)$ and $_2F_1(...)$ denote the Euler beta function and the Gauss hypergeometric function, respectively.
Setting $\mu=m+1,~\nu=1,~n=t,~r=k,~a=\rho,~c=-b,~d=a \eta$ in (\ref{A8}) and simplifying, we obtain the following explicit representation for the integral (\ref{A7}):
\begin{equation}\label{A9}
\mathcal{P}_{k,m,t}(a,b,\alpha,\rho)=
\frac{\rho^{m+1}}{(m+1)(a\eta)^{k}}\sum_{q=0}^t(-1)^{q+t}{t\choose q}\ln^q(a\eta)
~F\left(k,m,t-q;\frac{b \rho}{a \eta}\right),
\end{equation}
where we introduced the notation
\begin{equation}\label{A10}
F(k,m,n;x)=\lim_{r\rightarrow k}\frac{\partial^{n}}{\partial r^n}
~_2F_1\left(r,m+1;m+2;x\right).
\end{equation}
It is necessary to note that problems arise when it is necessary to calculate (directly) the Gauss hypergeometric function (and/or its derivatives in respect the first parameter) from Eq.(\ref{A10}) with an integer first parameter.
Thus, the next step is to obtain the trouble-free explicit representation for the function (\ref{A10}).
Using representation
\begin{equation}\label{A11}
_2F_1\left(r,m+1;m+2;x\right)=(m+1)x^{-(m+1)}B_x(m+1,1-r)
\end{equation}
for the hypergeometric function, and the integral representation
\begin{equation}\label{A12}
B_x(a,b)=\int_0^x t^{a-1}(1-t)^{b-1} dt
\end{equation}
for the incomplete beta function, we can give the following sequential derivation of the integral representation for the function (\ref{A10}):
\begin{eqnarray}\label{A13}
F(k,m,n;x)=(m+1)x^{-(m+1)}\lim_{r\rightarrow k}\int_0^x t^m \frac{\partial^{n}}{\partial r^n}(1-t)^{-r}dt=
\frac{(m+1)}{x^{m+1}}\int_0^x\frac{t^m[-\ln(1-t)]^n}{(1-t)^k}dt=
~~~\nonumber~\\
\frac{(m+1)}{x^{m+1}}\int_{1-x}^1y^{-k}(1-y)^m (-\ln y)^n~dy=
\frac{(m+1)}{x^{m+1}}\sum_{q=0}^m(-1)^q{m\choose q}\int_{1-x}^1 y^{q-k}(-\ln y)^n~dy.~~~~~~~~~~~
\end{eqnarray}
Using the explicit result of the last integration, we finally obtain:
\begin{eqnarray}\label{A13}
F(k,m,n;x)=\frac{(m+1)}{x^{m+1}}\sum_{q=0}^m(-1)^q{m\choose q}w_{k,n}(x;q),
\end{eqnarray}
where
\begin{equation}\label{A14}
w_{k,n}(x;q=k-1)=\frac{[-\ln(1-x)]^{n+1}}{n+1},
\end{equation}
\begin{equation}\label{A15}
w_{k,n}(x;q\neq k-1)= \frac{n!}{(q+1-k)^{n+1}}\left[1-(1-x)^{q+1-k}\sum_{s=0}^n\frac{(k-q-1)^s\ln^s(1-x)}{s!}\right].
\end{equation}
The result (\ref{A13})-(\ref{A15}) is valid for non-negative integer $m$ and $n$, real $x$ and integer $k$.

We have reduced the 3D integral (\ref{21}) to 1D integral (\ref{A3}).
It is useful to note that the accuracy and calculation time for the latter integral depend significantly on the choice of the method of integration. We used Wolfram Mathematica for all our calculations.
In particular, to calculate the integrals (\ref{A3}) we used the "GaussBerntsenEspelidRule" method, which significantly improves the accuracy and reduces the calculation time.
Gaussian quadrature uses optimal sampling points (through polynomial interpolation) to form a weighted sum of the integrand values over these points. On a subset of these sampling points a lower order quadrature rule can be made. The difference between the two rules can be used to estimate the error. Berntsen and Espelid derived error estimation rules by removing the central point of Gaussian rules with odd number of sampling points.

An additional remark concerns the behaviour of the integrand on the RHS of Eq. (\ref{A3}) as hyperspherical angle $\alpha$ approaches zero.
It is clear that $\eta=\xi=1$ for $\alpha=0$.
Therefore, it follows from representation (\ref{A6}) that the integrand on the RHS of Eq.(\ref{A3}) approaches zero as $\alpha\rightarrow 0$. The last statement is true at least for the cases $L=1,2$ (see Sections \ref{S2} and \ref{S3}) we are interested. The problem is that according to (\ref{A6}) this zero arises as the difference between two large but equal quantities. There are two solutions to the problem. The first is to apply high-precision calculations, at least in the vicinity of $\alpha=0$. The second is to use a series expansion for the RHS of Eq.(\ref{A6}) near $\alpha=0$.

\newpage

\begin{table}
\caption{Factor $ \widetilde{b}_{k'}$ and components of coefficients  $b_{k',n,m}=\widetilde{q}_{k',n,m}\sum_{j=0}^{J_{k'}}q_{k',n,m}^{(j)}\pi^j$~($0\leq k'\equiv k/2\leq 5$),
  \newline  representing the polynomials in $\eta$ and $\xi$  according to Eq.(\ref{15}).
  The upper limit of summation is defined as $J_{k'}=k'-2$ for $k'\geq2$ and $J_{k'}=0$ for $k'<2$.}
\begin{tabular}{|c|c|c|c|c|c|c|c|c|c|}
\hline  $k'$ &$\widetilde{b}_{k'}$&$N^{\underline{o}}$ &$n$& $m$ &$\widetilde{q}_{k',n,m}$& $q_{k',n,m}^{(0)}$& $q_{k',n,m}^{(1)}$& $q_{k',n,m}^{(2)}$& $q_{k',n,m}^{(3)}$ \\
\hline
\hline
 5&$(2-\pi)(5\pi-14)/$ & 1 & 0 & 0 & 1  & \sm{-1089387008240}& \sm{1047586705720}  &  \sm{-335763343914}  &  \sm{35868493935} \\
  & \sm{5012817232500000}   & 2 & 0 & 2 & 15 & \sm{4286509281040} & \sm{-4129722511880} &  \sm{1326134522774}  &\sm{-141939797505}  \\
  & & 3 & 0 & 4 & 60 &\sm{-5335564976800}&\sm{ 5141563833740} &\sm{-1651435721371} &\sm{ 176799081450}  \\
  & & 4 & 0 & 6 & 40 &\sm{12056179167440}&\sm{-11618550815440} &\sm{ 3732041258007}&\sm{-399570465060}  \\
  & & 5 & 0 & 8 &-15 &\sm{ 18776793358080}&\sm{-18095537797140}&\sm{ 5812646794643}&\sm{-622341848670}  \\
  & & 6 & 0 &10 &  3 &\sm{ 18776793358080}&\sm{-18095537797140}&\sm{ 5812646794643}&\sm{-622341848670}  \\
  & & 7 & 8 & 0 &  5 &\sm{-31900530538240}&\sm{ 30758808723020}&\sm{-9885507900549}&\sm{1058977460610} \\
  & & 8 & 8 & 2 & -5 &\sm{-31900530538240}&\sm{ 30758808723020}&\sm{-9885507900549}&\sm{1058977460610} \\
  & & 9 & 6 & 0 &-20 &\sm{-31900530538240}&\sm{ 30758808723020}&\sm{-9885507900549}&\sm{1058977460610} \\
  & &10 & 6 & 2 & 20 &\sm{-31900530538240}&\sm{ 30758808723020}&\sm{-9885507900549}&\sm{1058977460610} \\
  & &11 & 2 & 0 &160 &\sm{-967277327240}  &\sm{ 932324864590 } &\sm{-299527103049} &\sm{32074335000}   \\
  & &12 & 2 & 2 &120 &\sm{9531787759840}  &\sm{-9188929121180} &\sm{2952640515759} &\sm{-316236217230} \\
  & &13 & 2 & 4 & 60 &\sm{-24726253970560}&\sm{23837487905180} &\sm{-7659813135081}&\sm{820411311690 } \\
  & &14 & 2 & 6 &-20 &\sm{-24726253970560}&\sm{23837487905180} &\sm{-7659813135081}&\sm{820411311690 } \\
  & &15 & 4 & 0 & 20 &\sm{-28031421229280}&\sm{2702950926466}  &\sm{-8687399488353}&\sm{930680120610 } \\
  & &16 & 4 & 2 &160 &\sm{413145907340}   &\sm{-399002669935}  &\sm{128448294159}  &\sm{-13783601115 } \\
  & &17 & 4 & 4 & 30 &\sm{24726253970560} &\sm{-23837487905180}&\sm{7659813135081} &\sm{-820411311690} \\
  & &18 & 4 & 6 &-10 &\sm{24726253970560} &\sm{-23837487905180}&\sm{7659813135081} &\sm{-820411311690} \\
\hline
 4&$(\pi-2)(5\pi-14)/$ &1 & 0 & 0 &  12 & \sm{483896}   &  \sm{-308420}   & \sm{49095} &    \\
  &\sm{15431472000}    &2 & 2 & 0 & 288 & \sm{1201904}  &  \sm{-775505}   & \sm{125073} &   \\
  &                    &3 & 4 & 0 &   4 & \sm{48803648} & \sm{-31601320}  & \sm{5117187 } &  \\
  &                    &4 & 6 & 0 &  -4 & \sm{92072192} & \sm{-59519500 } & \sm{9619815 } & \\
  &                    &5 & 8 & 0 &   1 & \sm{92072192} & \sm{-59519500 } & \sm{9619815 } & \\
  &                    &6 & 0 & 2 & -96 & \sm{1685800}  & \sm{-1083925  } & \sm{174168 } & \\
  &                    &7 & 0 & 4 &  12 & \sm{37545376} & \sm{-24164696 } & \sm{3886989 } & \\
  &                    &8 & 0 & 6 & -12 & \sm{30802176} & \sm{-19828996 } & \sm{3190317 } & \\
  &                    &9 & 0 & 8 &   3 & \sm{30802176} & \sm{-19828996 } & \sm{3190317 } & \\
  &                    &10& 2 & 2 & -24 & \sm{46119680} & \sm{-29773780 } & \sm{4804833 } & \\
  &                    &11& 2 & 4 &  12 & \sm{46119680} & \sm{-29773780 } & \sm{4804833 } & \\
  &                    &12& 4 & 2 &  12 & \sm{46119680} & \sm{-29773780 } & \sm{4804833 } & \\
  &                    &13& 4 & 4 &  -6 & \sm{46119680} & \sm{-29773780 } & \sm{4804833 } & \\
  \hline
 3&$(2-\pi)(5\pi-14)/$ &1 & 0 & 0 & -2 & \sm{106}   &  \sm{-33}   &&   \\
  &\sm{170100}         &2 & 2 & 0 & -2 & \sm{2064}  &  \sm{-675}  &&   \\
  &                    &3 & 4 & 0 &  1 & \sm{2064}  &  \sm{-675}  &&   \\
  &                    &4 & 0 & 2 &  4 & \sm{609}   &  \sm{-195}  &&   \\
  &                    &5 & 0 & 4 & -3 & \sm{1112}  &  \sm{-357}  &&   \\
  &                    &6 & 0 & 6 &  1 & \sm{1112}  &  \sm{-357}  &&   \\
  &                    &7 & 2 & 2 &  2 & \sm{2064}  &  \sm{-675}  &&   \\
  &                    &8 & 4 & 2 & -1 & \sm{2064}  &  \sm{-675}  &&   \\
   \hline
   2&$(\pi-2)(5\pi-14)/$ &1 & 0 & 0 &   1 & \sm{1}   &&&   \\
  &\sm{180}              &2 & 2 & 0 &   1 & \sm{4}  &&&   \\
  &                      &3 & 4 & 0 &   1 & \sm{-2}  &&&   \\
  &                      &4 & 0 & 2 &  -1 & \sm{4}  &&&   \\
  &                      &5 & 0 & 4 &  -1 & \sm{-2}  &&&   \\
  \hline
   1&$-(\pi-2)/3$       &1 & 0 & 0 &   1 & \sm{1}   &&&   \\
  &              &2 & 0 & 2 &   1 & \sm{-1}  &&&   \\
  \hline
   0&$1$       &1 & 0 & 0 &   1 & \sm{1}   &&&   \\
  \hline
 \end{tabular}
\label{T1}
\end{table}

\begin{table}
\caption{Absolute values of the ground state energies $E$ and the corresponding FE coefficients $a_{21}$ calculated by VMNN with basis size $N_b$. The sixth column presents an estimate of the deviation from the \emph{virial} theorem. The two relevant matrix elements are presented in the last two columns.}
\begin{tabular}{|c|c|c|c|c|c|c|c|}
\hline atom/ion & $Z$ & $a_{21}$  & $N_b$ & $|E|$ & $\langle V \rangle/\langle T \rangle+2 $ &$\langle\delta(\textbf{r}_1)\rangle $ &$\langle\delta(\textbf{r}_1)\delta(\textbf{r}_2)\rangle $\\
\hline
\hline           &$1.5$ & 0.38238 & 517 &   1.465 279 051 825 74  & 4.5~(-13)&0.682115722 &0.20291\\
\hline He        & $2$  & 0.47675 & 517 &   2.903 724 377 034 119 &9.7~(-15)&1.8104293184&1.86874 \\
\hline Li$^+$    & $3$  & 0.62285 & 517 &   7.279 913 412 669 306 &6.8~(-16)&6.852009437&33.32118\\
\hline Be$^{2+}$ & $4$  & 0.72787 & 517 &  13.655 566 238 423 587 &2.9~(-16)&17.198172547&231.0389\\
\hline B$^{3+}$  & $5$  & 0.80232 & 517 &  22.030 971 580 242 781 &1.8~(-16)&34.75874366&996.0099\\
\hline C$^{4+}$  & $6$  & 0.85240 & 517 &  32.406 246 601 898 530 &1.2~(-16)&61.44357805&3221.850\\
\hline N$^{5+}$  & $7$  & 0.88221 & 442 &  44.781 445 148 772 703 & 5.2~(-16)&99.1625345&8596.074\\
\hline O$^{6+}$  & $8$  & 0.89466 & 442 &  59.156 595 122 757 924 & 3.6~(-16)&149.8254726&19974.31\\
\hline F$^{7+}$  & $9$  & 0.89193 & 442 &  75.531 712 363 959 490 & 2.6~(-16)&215.3422520& 41827.47\\
\hline Ne$^{8+}$ & $10$ & 0.87572 & 442 &  93.906 806 515 037 548 & 2.0~(-16)&297.6227308&80761.89\\
\hline Na$^{9+}$ & $11$ & 0.84738 & 442 & 114.281 883 776 072 721 & 1.5~(-16)&398.5767736&146112.4\\
\hline Mg$^{10+}$& $12$ & 0.80801 & 442 & 136.656 948 312 646 929 & 1.2~(-16)&520.1142308&250608.2\\
\hline Al$^{11+}$& $13$ & 0.75854 & 442 & 161.032 003 026 058 359 & 1.0~(-16)&664.1449752&411112.2\\
\hline Si$^{12+}$& $14$ & 0.69974 & 442 & 187.407 049 998 662 925 & 8.5~(-17)&832.5788588&649432.6\\
\hline P$^{13+}$ & $15$ & 0.63229 & 442 & 215.782 090 763 537 159 & 7.2~(-17)&1027.325721&993207.9\\
\hline S$^{14+}$ & $16$ & 0.55675 & 442 & 246.157 126 474 254 738 & 6.1~(-17)&1250.295442&147686.5~(+1)\\
\hline Cl$^{15+}$ & $17$& 0.47364 & 442 & 278.532 158 015 400 094 & 5.3~(-17)&1503.397895&214264.8~(+1)\\
\hline Ar$^{16+}$ & $18$& 0.38340 & 442 & 312.907 186 076 611 148 & 4.6~(-17)&1788.543432&304172.8~(+1)\\
\hline K$^{17+}$ & $19$ & 0.28642 & 442 & 349.282 211 203 453 166 & 4.1~(-17)&2107.640384&423536.9~(+1)\\
\hline Ca$^{18+}$& $20$ & 0.18307 & 442 & 387.657 233 833 158 555 & 3.7~(-17)&2462.600114&579618.9~(+1)\\
\hline Sc$^{19+}$& $21$ & 0.07366 & 442 & 428.032 254 320 234 690 & 3.2~(-17)&2855.332029&780947.6~(+1)\\
\hline Ti$^{20+}$& $22$ &-0.04152 & 442 & 470.407 272 955 138 383 & 2.9~(-17)&3287.745943&103745.9~(+2)\\
\hline V $^{21+}$& $23$ &-0.16220 & 442 & 514.782 289 978 111 773 & 2.6~(-17)&3761.751772&136064.1~(+2)\\
\hline Cr$^{22+}$& $24$ &-0.28816 & 442 & 561.157 305 589 581 271 & 2.4~(-17)&4279.259269&176369.1~(+2)\\
\hline Mn$^{23+}$& $25$ &-0.41916 & 442 & 609.532 319 958 075 745 & 2.2~(-17)&4842.178389&226167.5~(+2)\\
\hline Fe$^{24+}$& $26$ &-0.55501 & 442 & 659.907 333 226 327 804 & 2.0~(-17)&5452.418955&287169.7~(+2)\\
\hline Co$^{25+}$& $27$ &-0.69552 & 442 & 712.282 345 516 026 550 & 1.8~(-17)&6111.890859&361307.4~(+2)\\
\hline Ni$^{26+}$& $28$ &-0.84052 & 442 & 766.657 356 931 557 099 & 1.7~(-17)&6822.503872&450752.1~(+2)\\
\hline Zn$^{28+}$& $30$ &-1.14334 & 442 & 881.407 377 488 360 605 & 1.5~(-17)&8404.792902&685562.3~(+2)\\
\hline Zr$^{38+}$& $40$ &-2.88133 & 442 & 1575.157 449 525 559 & 7.7~(-18)&20034.060&323922.3~(+3)\\
\hline Sn$^{48+}$& $50$ &-4.92537 & 442 & 2468.907 492 812 712 & 4.7~(-18)&39260.262&151406.5~(+4)\\
\hline Nd$^{58+}$& $60$ &-7.21331 & 442 & 3562.657 521 679 319 & 3.2~(-18)&67993.257&455483.1~(+4)\\
\hline Yb$^{68+}$& $70$ &-9.70406 & 442 & 4856.407 542 342 006 & 2.3~(-18)&108142.90&115468.8~(+5)\\
\hline Hg$^{78+}$& $80$ &-12.3685 & 442 & 6350.157 557 832 474 & 1.7~(-18)&161619.06&258314.5~(+5)\\
\hline Th$^{88+}$& $90$ &-15.1846 & 442 & 8043.907 569 884 711 & 1.3~(-18)&230331.59&525303.7~(+5)\\
\hline Fm$^{98+}$& $100$&-18.1357 & 442 & 9937.657 579 529 067 & 1.1~(-18)&316190.36&990905.1~(+5)\\
 \hline
\end{tabular}
\label{T2}
\end{table}

\end{document}